

%
\catcode`@=11
%
%
%
\def\lsim{\mathchoice
  {\mathrel{\lower.8ex\hbox{$\displaystyle\buildrel<\over\sim$}}}
  {\mathrel{\lower.8ex\hbox{$\textstyle\buildrel<\over\sim$}}}
  {\mathrel{\lower.8ex\hbox{$\scriptstyle\buildrel<\over\sim$}}}
  {\mathrel{\lower.8ex\hbox{$\scriptscriptstyle\buildrel<\over\sim$}}} }
\def\gsim{\mathchoice
  {\mathrel{\lower.8ex\hbox{$\displaystyle\buildrel>\over\sim$}}}
  {\mathrel{\lower.8ex\hbox{$\textstyle\buildrel>\over\sim$}}}
  {\mathrel{\lower.8ex\hbox{$\scriptstyle\buildrel>\over\sim$}}}
  {\mathrel{\lower.8ex\hbox{$\scriptscriptstyle\buildrel>\over\sim$}}} }
%
%
%
%
\def\quad@rato#1#2{{\vcenter{\vbox{
        \hrule height#2pt
        \hbox{\vrule width#2pt height#1pt \kern#1pt \vrule width#2pt}
        \hrule height#2pt} }}}
\def\quadratello{\mathchoice
\quad@rato5{.5}\quad@rato5{.5}\quad@rato{3.5}{.35}\quad@rato{2.5}{.25} }
%
%
\font\s@a=cmss10\font\s@b=cmss8
\def\reali{{\mathchoice
 {\hbox{\s@a l\kern-.5mm R}}
 {\hbox{\s@a l\kern-.5mm R}}
 {\hbox{\s@b l\kern-.4mm R}}
 {\hbox{\s@b l\kern-.4mm R}}  }}
\def\naturali{{\mathchoice
 {\hbox{\s@a l\kern-.5mm N}}
 {\hbox{\s@a l\kern-.5mm N}}
 {\hbox{\s@b l\kern-.4mm N}}
 {\hbox{\s@b l\kern-.4mm N}}  }}
\def\interi{{\mathchoice
 {\hbox{\s@a Z\kern-1.5mm Z}}
 {\hbox{\s@a Z\kern-1.5mm Z}}
 {\hbox{{\s@b Z\kern-1.2mm Z}}}
 {\hbox{{\s@b Z\kern-1.2mm Z}}}  }}
\def\complessi{{\mathchoice
 {\hbox{\s@a C\kern-1.4mm\raise.2mm\hbox{\s@b l}\kern.8mm}}
 {\hbox{\s@a C\kern-1.4mm\raise.2mm\hbox{\s@b l}\kern.8mm}}
 {\hbox{\s@b C\kern-1.1mm\raise.13mm\hbox{\s@b l}\kern.5mm}}
 {\hbox{\s@b C\kern-1.1mm\raise.13mm\hbox{\s@b l}\kern.5mm}} }}
\def\toro{{\mathchoice
 {\hbox{\s@a T\kern-1.9mm T}}
 {\hbox{\s@a T\kern-1.9mm T}}
 {\hbox{\s@b T\kern-1.4mm T}}
 {\hbox{\s@b T\kern-1.4mm T}}  }}
\def\unity{{\hbox{\s@a 1\kern-.8mm l}}}
%
%
\def\footreali{\hbox{\s@b l\kern-.4mm R}}
\def\footnaturali{\hbox{\s@b l\kern-.4mm N}}
\def\footinteri{\hbox{{\s@b Z\kern-1.2mm Z}}}
\def\footcomplessi{\hbox{\s@b C\kern-1.1mm\raise.13mm\hbox{\s@b l}\kern.5mm}}
\def\foottoro{\hbox{\s@b T\kern-1.4mm T}}

%
%
\font\bold@mit=cmmib10
\def\setbmit{\textfont1=\bold@mit}
\def\bmit#1{\hbox{\textfont1=\bold@mit$#1$}}
%
\catcode`@=12

%
%
\catcode`@=11 
%
%

\def\b@lank{ }

\newif\if@simboli
\newif\if@riferimenti
\newif\if@bozze

\def\bozze{\@bozzetrue\font\tt@bozze=cmtt8}
\def\og@gi{\number\day\space\ifcase\month\or 
   gennaio\or febbraio\or marzo\or aprile\or maggio\or giugno\or 
   luglio\or agosto\or settembre\or ottobre\or novembre\or dicembre\fi
   \space\number\year}
\newcount\min@uti
\newcount\or@a
\newcount\ausil@iario
\min@uti=\number\time
\or@a=\number\time
\divide\or@a by 60
\ausil@iario=-\number\or@a
\multiply\ausil@iario by 60
\advance\min@uti by \number\ausil@iario
\def\ora@esecuzione{\the\or@a:\the\min@uti}  
\def\makefootline{\baselineskip=24pt\line{\the\footline}
    \if@bozze\vskip-10pt\tt@bozze
             \noindent \jobname\hfill\og@gi, ore \ora@esecuzione\fi}

\newwrite\file@simboli
\def\simboli{
    \immediate\write16{ !!! Genera il file \jobname.SMB }
    \@simbolitrue\immediate\openout\file@simboli=\jobname.smb}

\newwrite\file@ausiliario
\def\riferimentifuturi{
    \immediate\write16{ !!! Genera il file \jobname.AUX }
    \@riferimentitrue\openin1 \jobname.aux
    \ifeof1\relax\else\closein1\relax\input\jobname.aux\fi
    \immediate\openout\file@ausiliario=\jobname.aux}

\newcount\eq@num\global\eq@num=0
\newcount\sect@num\global\sect@num=0

\newif\if@ndoppia
\def\numerazionedoppia{\@ndoppiatrue\gdef\la@sezionecorrente{\the\sect@num}}

\def\se@indefinito#1{\expandafter\ifx\csname#1\endcsname\relax}
\def\spo@glia#1>{} 

\newif\if@primasezione
\@primasezionetrue

\def\s@ection#1\par{\immediate
    \write16{#1}\if@primasezione\global\@primasezionefalse\else\goodbreak
    \vskip\spaziosoprasez\fi\noindent
    {\sezfont #1}\nobreak\vskip\spaziosottosez\nobreak\noindent}
%

\font\sezfont=cmbx10

\def\sezpreset#1{\global\sect@num=#1
    \immediate\write16{ !!! sez-preset = #1 }   }

\def\spaziosoprasez{30pt plus 20pt}
\def\spaziosottosez{10pt}
\def\spaziotitsez{3truemm}

\def\sref#1{\se@indefinito{@s@#1}\immediate\write16{ ??? \string\sref{#1}
    non definita !!!}
    \expandafter\xdef\csname @s@#1\endcsname{??}\fi\csname @s@#1\endcsname}

\def\autosez#1#2\par{
    \global\advance\sect@num by 1\if@ndoppia\global\eq@num=0\fi
    \xdef\la@sezionecorrente{\the\sect@num}
    \def\usa@getta{1}\se@indefinito{@s@#1}\def\usa@getta{2}\fi
    \expandafter\ifx\csname @s@#1\endcsname\la@sezionecorrente\def
    \usa@getta{2}\fi
    \ifodd\usa@getta\immediate\write16
      { ??? possibili riferimenti errati a \string\sref{#1} !!!}\fi
    \expandafter\xdef\csname @s@#1\endcsname{\la@sezionecorrente}
    \immediate\write16{\la@sezionecorrente. #2}
    \if@simboli
      \immediate\write\file@simboli{ }\immediate\write\file@simboli{ }
      \immediate\write\file@simboli{  Sezione 
                                  \la@sezionecorrente :   sref.   #1}
      \immediate\write\file@simboli{ } \fi
    \if@riferimenti
      \immediate\write\file@ausiliario{\string\expandafter\string\edef
      \string\csname\b@lank @s@#1\string\endcsname{\la@sezionecorrente}}\fi
    \goodbreak\vskip\spaziosoprasez
    \noindent\if@bozze\llap{\tt@bozze#1\ }\fi
      {\sezfont\the\sect@num.\hskip\spaziotitsez #2}\par\nobreak
    \vskip\spaziosottosez\nobreak\noindent}

\def\semiautosez#1#2\par{
    \gdef\la@sezionecorrente{#1}\if@ndoppia\global\eq@num=0\fi
    \if@simboli
      \immediate\write\file@simboli{ }\immediate\write\file@simboli{ }
      \immediate\write\file@simboli{  Sezione ** : sref.
          \expandafter\spo@glia\meaning\la@sezionecorrente}
      \immediate\write\file@simboli{ }\fi
    \s@ection#2\par}


\def\eqpreset#1{\global\eq@num=#1
     \immediate\write16{ !!! eq-preset = #1 }     }

\def\eqref#1{\se@indefinito{@eq@#1}
    \immediate\write16{ ??? \string\eqref{#1} non definita !!!}
    \expandafter\xdef\csname @eq@#1\endcsname{??}
    \fi\csname @eq@#1\endcsname}

\def\eqlabel#1{\global\advance\eq@num by 1
    \if@ndoppia\xdef\il@numero{\la@sezionecorrente.\the\eq@num}
       \else\xdef\il@numero{\the\eq@num}\fi
    \def\usa@getta{1}\se@indefinito{@eq@#1}\def\usa@getta{2}\fi
    \expandafter\ifx\csname @eq@#1\endcsname\il@numero\def\usa@getta{2}\fi
    \ifodd\usa@getta\immediate\write16
       { ??? possibili riferimenti errati a \string\eqref{#1} !!!}\fi
    \expandafter\xdef\csname @eq@#1\endcsname{\il@numero}
    \if@ndoppia
       \def\usa@getta{\expandafter\spo@glia\meaning
       \la@sezionecorrente.\the\eq@num}
       \else\def\usa@getta{\the\eq@num}\fi
    \if@simboli
       \immediate\write\file@simboli{  Equazione 
            \usa@getta :  eqref.   #1}\fi
    \if@riferimenti
       \immediate\write\file@ausiliario{\string\expandafter\string\edef
       \string\csname\b@lank @eq@#1\string\endcsname{\usa@getta}}\fi}

\def\autoreqno#1{\eqlabel{#1}\eqno(\csname @eq@#1\endcsname)
       \if@bozze\rlap{\tt@bozze\ #1}\fi}
\def\autoleqno#1{\eqlabel{#1}\leqno\if@bozze\llap{\tt@bozze#1\ }
       \fi(\csname @eq@#1\endcsname)}
\def\eqrefp#1{(\eqref{#1})}
\def\numeriadestra{\let\autoeqno=\autoreqno}
\def\numaeriasinistra{\let\autoeqno=\autoleqno}
\numeriadestra

\newcount\cit@num\global\cit@num=0

\newwrite\file@bibliografia
\newif\if@bibliografia
\@bibliografiafalse

\def\lp@cite{[}
\def\rp@cite{]}
\def\trap@cite#1{\lp@cite #1\rp@cite}
\def\lp@bibl{[}
\def\rp@bibl{]}
\def\trap@bibl#1{\lp@bibl #1\rp@bibl}

\def\refe@renza#1{\if@bibliografia\immediate        
    \write\file@bibliografia{
    \string\item{\trap@bibl{\cref{#1}}}\string
    \bibl@ref{#1}\string\bibl@skip}\fi}

\def\ref@ridefinita#1{\if@bibliografia\immediate\write\file@bibliografia{ 
    \string\item{?? \trap@bibl{\cref{#1}}} ??? tentativo di ridefinire la 
      citazione #1 !!! \string\bibl@skip}\fi}

\def\bibl@ref#1{\se@indefinito{@ref@#1}\immediate
    \write16{ ??? biblitem #1 indefinito !!!}\expandafter\xdef
    \csname @ref@#1\endcsname{ ??}\fi\csname @ref@#1\endcsname}

\def\c@label#1{\global\advance\cit@num by 1\xdef            
   \la@citazione{\the\cit@num}\expandafter
   \xdef\csname @c@#1\endcsname{\la@citazione}}

\def\bibl@skip{\vskip 0truept}


\def\stileincite#1#2{\global\def\lp@cite{#1}\global
    \def\rp@cite{#2}}
\def\stileinbibl#1#2{\global\def\lp@bibl{#1}\global
    \def\rp@bibl{#2}}

\def\citpreset#1{\global\cit@num=#1
    \immediate\write16{ !!! cit-preset = #1 }    }

\def\autobibliografia{\global\@bibliografiatrue\immediate
    \write16{ !!! Genera il file \jobname.BIB}\immediate
    \openout\file@bibliografia=\jobname.bib}

\def\cref#1{\se@indefinito                  
   {@c@#1}\c@label{#1}\refe@renza{#1}\fi\csname @c@#1\endcsname}

\def\cite#1{\trap@cite{\cref{#1}}}                        
\def\upcite#1{$^{\,\trap@cite{\cref{#1}}}$}               
\def\ccite#1#2{\trap@cite{\cref{#1},\cref{#2}}}           
\def\upccite#1#2{$^{\,\trap@cite{\cref{#1},\cref{#2}}}$}  
\def\cccite#1#2#3{\trap@cite{\cref{#1},\cref{#2},\cref{#3}}}          
\def\upcccite#1#2#3{$^{\,\trap@cite{\cref{#1},\cref{#2},\cref{#3}}}$} 
\def\ncite#1#2{\trap@cite{\cref{#1}--\cref{#2}}}          
\def\upncite#1#2{$^{\,\trap@cite{\cref{#1}-\cref{#2}}}$}  

\def\clabel#1{\se@indefinito{@c@#1}\c@label           
    {#1}\refe@renza{#1}\else\c@label{#1}\ref@ridefinita{#1}\fi}

\def\biblskip#1{\def\bibl@skip{\vskip #1}}           

\def\insertbibliografia{\if@bibliografia             
    \immediate\write\file@bibliografia{ }
    \immediate\closeout\file@bibliografia
    \catcode`@=11\input\jobname.bib\catcode`@=12\fi}


\def\commento#1{\relax} 
\def\biblitem#1#2\par{\expandafter\xdef\csname @ref@#1\endcsname{#2}}


\catcode`@=12

%
%
%
%
%
%
%
%

\ifx\undefined\psfig\else\endinput\fi

%
\edef\psfigRestoreAt{\catcode`@=\number\catcode`@\relax}
\catcode`\@=11\relax
\newwrite\@unused
\def\typeout#1{{\let\protect\string\immediate\write\@unused{#1}}}
\typeout{psfig/tex 1.6b}


\def\figurepath{./}

%
%
\def\@nnil{\@nil}
\def\@empty{}
\def\@psdonoop#1\@@#2#3{}
\def\@psdo#1:=#2\do#3{\edef\@psdotmp{#2}\ifx\@psdotmp\@empty \else
    \expandafter\@psdoloop#2,\@nil,\@nil\@@#1{#3}\fi}
\def\@psdoloop#1,#2,#3\@@#4#5{\def#4{#1}\ifx #4\@nnil \else
       #5\def#4{#2}\ifx #4\@nnil \else#5\@ipsdoloop #3\@@#4{#5}\fi\fi}
\def\@ipsdoloop#1,#2\@@#3#4{\def#3{#1}\ifx #3\@nnil 
       \let\@nextwhile=\@psdonoop \else
      #4\relax\let\@nextwhile=\@ipsdoloop\fi\@nextwhile#2\@@#3{#4}}
\def\@tpsdo#1:=#2\do#3{\xdef\@psdotmp{#2}\ifx\@psdotmp\@empty \else
    \@tpsdoloop#2\@nil\@nil\@@#1{#3}\fi}
\def\@tpsdoloop#1#2\@@#3#4{\def#3{#1}\ifx #3\@nnil 
       \let\@nextwhile=\@psdonoop \else
      #4\relax\let\@nextwhile=\@tpsdoloop\fi\@nextwhile#2\@@#3{#4}}
%
%
%
\newread\ps@stream
\newif\ifnot@eof       
\newif\if@noisy        
\newif\if@atend        
\newif\if@psfile       
%
%
{\catcode`\%=12\global\gdef\epsf@start{
\def\epsf@PS{PS}
\def\epsf@getbb#1{%
%
%
\openin\ps@stream=#1
\ifeof\ps@stream\typeout{Error, File #1 not found}\else
%
%
   {\not@eoftrue \chardef\other=12
    \def\do##1{\catcode`##1=\other}\dospecials \catcode`\ =10
    \loop
       \if@psfile
	  \read\ps@stream to \epsf@fileline
       \else{
	  \obeyspaces
          \read\ps@stream to \epsf@tmp\global\let\epsf@fileline\epsf@tmp}
       \fi
       \ifeof\ps@stream\not@eoffalse\else
%
%
       \if@psfile\else
       \expandafter\epsf@test\epsf@fileline:. \\%
       \fi
%
%
          \expandafter\epsf@aux\epsf@fileline:. \\%
       \fi
   \ifnot@eof\repeat
   }\closein\ps@stream\fi}%
%
%
\long\def\epsf@test#1#2#3:#4\\{\def\epsf@testit{#1#2}
			\ifx\epsf@testit\epsf@start\else
\typeout{Warning! File does not start with `\epsf@start'.  It may not be a PostScript file.}
			\fi
			\@psfiletrue} 
%
%
{\catcode`\%=12\global\let\epsf@percent=
%
%
%
\long\def\epsf@aux#1#2:#3\\{\ifx#1\epsf@percent
   \def\epsf@testit{#2}\ifx\epsf@testit\epsf@bblit
	\@atendfalse
        \epsf@atend #3 . \\%
	\if@atend	
	   \if@verbose{
		\typeout{psfig: found `(atend)'; continuing search}
	   }\fi
        \else
        \epsf@grab #3 . . . \\%
        \not@eoffalse
        \global\no@bbfalse
        \fi
   \fi\fi}%
%
%
\def\epsf@grab #1 #2 #3 #4 #5\\{%
   \global\def\epsf@llx{#1}\ifx\epsf@llx\empty
      \epsf@grab #2 #3 #4 #5 .\\\else
   \global\def\epsf@lly{#2}%
   \global\def\epsf@urx{#3}\global\def\epsf@ury{#4}\fi}%
%
%
\def\epsf@atendlit{(atend)} 
\def\epsf@atend #1 #2 #3\\{%
   \def\epsf@tmp{#1}\ifx\epsf@tmp\empty
      \epsf@atend #2 #3 .\\\else
   \ifx\epsf@tmp\epsf@atendlit\@atendtrue\fi\fi}


\chardef\letter = 11
\chardef\other = 12

\newif \ifdebug 
\newif\ifc@mpute 
\c@mputetrue 

\let\then = \relax
\def\r@dian{pt }
\let\r@dians = \r@dian
\let\dimensionless@nit = \r@dian
\let\dimensionless@nits = \dimensionless@nit
\def\internal@nit{sp }
\let\internal@nits = \internal@nit
\newif\ifstillc@nverging
\def \Mess@ge #1{\ifdebug \then \message {#1} \fi}

{ 
	\catcode `\@ = \letter
	\gdef \nodimen {\expandafter \n@dimen \the \dimen}
	\gdef \term #1 #2 #3%
	       {\edef \t@ {\the #1}
		\edef \t@@ {\expandafter \n@dimen \the #2\r@dian}%
		\t@rm {\t@} {\t@@} {#3}%
	       }
	\gdef \t@rm #1 #2 #3%
	       {{%
		\count 0 = 0
		\dimen 0 = 1 \dimensionless@nit
		\dimen 2 = #2\relax
		\Mess@ge {Calculating term #1 of \nodimen 2}%
		\loop
		\ifnum	\count 0 < #1
		\then	\advance \count 0 by 1
			\Mess@ge {Iteration \the \count 0 \space}%
			\Multiply \dimen 0 by {\dimen 2}%
			\Mess@ge {After multiplication, term = \nodimen 0}%
			\Divide \dimen 0 by {\count 0}%
			\Mess@ge {After division, term = \nodimen 0}%
		\repeat
		\Mess@ge {Final value for term #1 of 
				\nodimen 2 \space is \nodimen 0}%
		\xdef \Term {#3 = \nodimen 0 \r@dians}%
		\aftergroup \Term
	       }}
	\catcode `\p = \other
	\catcode `\t = \other
	\gdef \n@dimen #1pt{#1} 
}

\def \Divide #1by #2{\divide #1 by #2} 

\def \Multiply #1by #2
       {{
	\count 0 = #1\relax
	\count 2 = #2\relax
	\count 4 = 65536
	\Mess@ge {Before scaling, count 0 = \the \count 0 \space and
			count 2 = \the \count 2}%
	\ifnum	\count 0 > 32767 
	\then	\divide \count 0 by 4
		\divide \count 4 by 4
	\else	\ifnum	\count 0 < -32767
		\then	\divide \count 0 by 4
			\divide \count 4 by 4
		\else
		\fi
	\fi
	\ifnum	\count 2 > 32767 
	\then	\divide \count 2 by 4
		\divide \count 4 by 4
	\else	\ifnum	\count 2 < -32767
		\then	\divide \count 2 by 4
			\divide \count 4 by 4
		\else
		\fi
	\fi
	\multiply \count 0 by \count 2
	\divide \count 0 by \count 4
	\xdef \product {#1 = \the \count 0 \internal@nits}%
	\aftergroup \product
       }}

\def\r@duce{\ifdim\dimen0 > 90\r@dian \then   
		\multiply\dimen0 by -1
		\advance\dimen0 by 180\r@dian
		\r@duce
	    \else \ifdim\dimen0 < -90\r@dian \then  
		\advance\dimen0 by 360\r@dian
		\r@duce
		\fi
	    \fi}

\def\Sine#1%
       {{%
	\dimen 0 = #1 \r@dian
	\r@duce
	\ifdim\dimen0 = -90\r@dian \then
	   \dimen4 = -1\r@dian
	   \c@mputefalse
	\fi
	\ifdim\dimen0 = 90\r@dian \then
	   \dimen4 = 1\r@dian
	   \c@mputefalse
	\fi
	\ifdim\dimen0 = 0\r@dian \then
	   \dimen4 = 0\r@dian
	   \c@mputefalse
	\fi
	\ifc@mpute \then
		\divide\dimen0 by 180
		\dimen0=3.141592654\dimen0
		\dimen 2 = 3.1415926535897963\r@dian 
		\divide\dimen 2 by 2 
		\Mess@ge {Sin: calculating Sin of \nodimen 0}%
		\count 0 = 1 
		\dimen 2 = 1 \r@dian 
		\dimen 4 = 0 \r@dian 
		\loop
			\ifnum	\dimen 2 = 0 
			\then	\stillc@nvergingfalse 
			\else	\stillc@nvergingtrue
			\fi
			\ifstillc@nverging 
			\then	\term {\count 0} {\dimen 0} {\dimen 2}%
				\advance \count 0 by 2
				\count 2 = \count 0
				\divide \count 2 by 2
				\ifodd	\count 2 
				\then	\advance \dimen 4 by \dimen 2
				\else	\advance \dimen 4 by -\dimen 2
				\fi
		\repeat
	\fi		
			\xdef \sine {\nodimen 4}%
       }}

\def\Cosine#1{\ifx\sine\UnDefined\edef\Savesine{\relax}\else
		             \edef\Savesine{\sine}\fi
	{\dimen0=#1\r@dian\advance\dimen0 by 90\r@dian
	 \Sine{\nodimen 0}
	 \xdef\cosine{\sine}
	 \xdef\sine{\Savesine}}}	      

\def\psdraft{
	\def\@psdraft{0}
}
\def\psfull{
	\def\@psdraft{100}
}

\psfull

\newif\if@draftbox
\def\psnodraftbox{
	\@draftboxfalse
}
\def\psdraftbox{
	\@draftboxtrue
}
\@draftboxtrue

\newif\if@prologfile
\newif\if@postlogfile
\def\pssilent{
	\@noisyfalse
}
\def\psnoisy{
	\@noisytrue
}
\psnoisy
\newif\if@bbllx
\newif\if@bblly
\newif\if@bburx
\newif\if@bbury
\newif\if@height
\newif\if@width
\newif\if@rheight
\newif\if@rwidth
\newif\if@angle
\newif\if@clip
\newif\if@verbose
\def\@p@@sclip#1{\@cliptrue}


\def\@p@@sfile#1{\def\@p@sfile{null}%
	        \openin1=#1
		\ifeof1\closein1%
		       \openin1=\figurepath#1
			\ifeof1\typeout{Error, File #1 not found}
			   \if@bbllx\if@bblly\if@bburx\if@bbury
			      \def\@p@sfile{#1}%
			   \fi\fi\fi\fi
			\else\closein1
			    \edef\@p@sfile{\figurepath#1}%
                        \fi%
		 \else\closein1%
		       \def\@p@sfile{#1}%
		 \fi}
\def\@p@@sfigure#1{\def\@p@sfile{null}%
	        \openin1=#1
		\ifeof1\closein1%
		       \openin1=\figurepath#1
			\ifeof1\typeout{Error, File #1 not found}
			   \if@bbllx\if@bblly\if@bburx\if@bbury
			      \def\@p@sfile{#1}%
			   \fi\fi\fi\fi
			\else\closein1
			    \def\@p@sfile{\figurepath#1}%
                        \fi%
		 \else\closein1%
		       \def\@p@sfile{#1}%
		 \fi}

\def\@p@@sbbllx#1{
		\@bbllxtrue
		\dimen100=#1
		\edef\@p@sbbllx{\number\dimen100}
}
\def\@p@@sbblly#1{
		\@bbllytrue
		\dimen100=#1
		\edef\@p@sbblly{\number\dimen100}
}
\def\@p@@sbburx#1{
		\@bburxtrue
		\dimen100=#1
		\edef\@p@sbburx{\number\dimen100}
}
\def\@p@@sbbury#1{
		\@bburytrue
		\dimen100=#1
		\edef\@p@sbbury{\number\dimen100}
}
\def\@p@@sheight#1{
		\@heighttrue
		\dimen100=#1
   		\edef\@p@sheight{\number\dimen100}
}
\def\@p@@swidth#1{
		\@widthtrue
		\dimen100=#1
		\edef\@p@swidth{\number\dimen100}
}
\def\@p@@srheight#1{
		\@rheighttrue
		\dimen100=#1
		\edef\@p@srheight{\number\dimen100}
}
\def\@p@@srwidth#1{
		\@rwidthtrue
		\dimen100=#1
		\edef\@p@srwidth{\number\dimen100}
}
\def\@p@@sangle#1{
		\@angletrue
		\edef\@p@sangle{#1} 
}
\def\@p@@ssilent#1{ 
		\@verbosefalse
}
\def\@p@@sprolog#1{\@prologfiletrue\def\@prologfileval{#1}}
\def\@p@@spostlog#1{\@postlogfiletrue\def\@postlogfileval{#1}}
\def\@cs@name#1{\csname #1\endcsname}
\def\@setparms#1=#2,{\@cs@name{@p@@s#1}{#2}}
%
%
\def\ps@init@parms{
		\@bbllxfalse \@bbllyfalse
		\@bburxfalse \@bburyfalse
		\@heightfalse \@widthfalse
		\@rheightfalse \@rwidthfalse
		\def\@p@sbbllx{}\def\@p@sbblly{}
		\def\@p@sbburx{}\def\@p@sbbury{}
		\def\@p@sheight{}\def\@p@swidth{}
		\def\@p@srheight{}\def\@p@srwidth{}
		\def\@p@sangle{0}
		\def\@p@sfile{}
		\def\@p@scost{10}
		\def\@sc{}
		\@prologfilefalse
		\@postlogfilefalse
		\@clipfalse
		\if@noisy
			\@verbosetrue
		\else
			\@verbosefalse
		\fi
}
%
%
\def\parse@ps@parms#1{
	 	\@psdo\@psfiga:=#1\do
		   {\expandafter\@setparms\@psfiga,}}
%
%
\newif\ifno@bb
\def\bb@missing{
	\if@verbose{
		\typeout{psfig: searching \@p@sfile \space  for bounding box}
	}\fi
	\no@bbtrue
	\epsf@getbb{\@p@sfile}
        \ifno@bb \else \bb@cull\epsf@llx\epsf@lly\epsf@urx\epsf@ury\fi
}	
\def\bb@cull#1#2#3#4{
	\dimen100=#1 bp\edef\@p@sbbllx{\number\dimen100}
	\dimen100=#2 bp\edef\@p@sbblly{\number\dimen100}
	\dimen100=#3 bp\edef\@p@sbburx{\number\dimen100}
	\dimen100=#4 bp\edef\@p@sbbury{\number\dimen100}
	\no@bbfalse
}
\newdimen\p@intvaluex
\newdimen\p@intvaluey
\def\rotate@#1#2{{\dimen0=#1 sp\dimen1=#2 sp
		  \global\p@intvaluex=\cosine\dimen0
		  \dimen3=\sine\dimen1
		  \global\advance\p@intvaluex by -\dimen3
		  \global\p@intvaluey=\sine\dimen0
		  \dimen3=\cosine\dimen1
		  \global\advance\p@intvaluey by \dimen3
		  }}
%
\def\compute@bb{
		\no@bbfalse
		\if@bbllx \else \no@bbtrue \fi
		\if@bblly \else \no@bbtrue \fi
		\if@bburx \else \no@bbtrue \fi
		\if@bbury \else \no@bbtrue \fi
		\ifno@bb \bb@missing \fi
		\ifno@bb \typeout{FATAL ERROR: no bb supplied or found}
			\no-bb-error
		\fi
		%
		\if@angle 
			\Sine{\@p@sangle}\Cosine{\@p@sangle}
	        	{\dimen100=\maxdimen\xdef\r@p@sbbllx{\number\dimen100}
					    \xdef\r@p@sbblly{\number\dimen100}
			                    \xdef\r@p@sbburx{-\number\dimen100}
					    \xdef\r@p@sbbury{-\number\dimen100}}
%
                        \def\minmaxtest{
			   \ifnum\number\p@intvaluex<\r@p@sbbllx
			      \xdef\r@p@sbbllx{\number\p@intvaluex}\fi
			   \ifnum\number\p@intvaluex>\r@p@sbburx
			      \xdef\r@p@sbburx{\number\p@intvaluex}\fi
			   \ifnum\number\p@intvaluey<\r@p@sbblly
			      \xdef\r@p@sbblly{\number\p@intvaluey}\fi
			   \ifnum\number\p@intvaluey>\r@p@sbbury
			      \xdef\r@p@sbbury{\number\p@intvaluey}\fi
			   }
			\rotate@{\@p@sbbllx}{\@p@sbblly}
			\minmaxtest
			\rotate@{\@p@sbbllx}{\@p@sbbury}
			\minmaxtest
			\rotate@{\@p@sbburx}{\@p@sbblly}
			\minmaxtest
			\rotate@{\@p@sbburx}{\@p@sbbury}
			\minmaxtest
			\edef\@p@sbbllx{\r@p@sbbllx}\edef\@p@sbblly{\r@p@sbblly}
			\edef\@p@sbburx{\r@p@sbburx}\edef\@p@sbbury{\r@p@sbbury}
		\fi
		\count203=\@p@sbburx
		\count204=\@p@sbbury
		\advance\count203 by -\@p@sbbllx
		\advance\count204 by -\@p@sbblly
		\edef\@bbw{\number\count203}
		\edef\@bbh{\number\count204}
}
%
%
\def\in@hundreds#1#2#3{\count240=#2 \count241=#3
		     \count100=\count240	
		     \divide\count100 by \count241
		     \count101=\count100
		     \multiply\count101 by \count241
		     \advance\count240 by -\count101
		     \multiply\count240 by 10
		     \count101=\count240	
		     \divide\count101 by \count241
		     \count102=\count101
		     \multiply\count102 by \count241
		     \advance\count240 by -\count102
		     \multiply\count240 by 10
		     \count102=\count240	
		     \divide\count102 by \count241
		     \count200=#1\count205=0
		     \count201=\count200
			\multiply\count201 by \count100
		 	\advance\count205 by \count201
		     \count201=\count200
			\divide\count201 by 10
			\multiply\count201 by \count101
			\advance\count205 by \count201
		     \count201=\count200
			\divide\count201 by 100
			\multiply\count201 by \count102
			\advance\count205 by \count201
		     \edef\@result{\number\count205}
}
\def\compute@wfromh{
		\in@hundreds{\@p@sheight}{\@bbw}{\@bbh}
		\edef\@p@swidth{\@result}
}
\def\compute@hfromw{
	        \in@hundreds{\@p@swidth}{\@bbh}{\@bbw}
		\edef\@p@sheight{\@result}
}
\def\compute@handw{
		\if@height 
			\if@width
			\else
				\compute@wfromh
			\fi
		\else 
			\if@width
				\compute@hfromw
			\else
				\edef\@p@sheight{\@bbh}
				\edef\@p@swidth{\@bbw}
			\fi
		\fi
}
\def\compute@resv{
		\if@rheight \else \edef\@p@srheight{\@p@sheight} \fi
		\if@rwidth \else \edef\@p@srwidth{\@p@swidth} \fi
}
%
\def\compute@sizes{
	\compute@bb
	\compute@handw
	\compute@resv
}
%
%
\def\psfig#1{\vbox {
	%
	\ps@init@parms
	\parse@ps@parms{#1}
	\compute@sizes
	\ifnum\@p@scost<\@psdraft{
		\if@verbose{
			\typeout{psfig: including \@p@sfile \space }
		}\fi
		\special{ps::[begin] 	\@p@swidth \space \@p@sheight \space
				\@p@sbbllx \space \@p@sbblly \space
				\@p@sbburx \space \@p@sbbury \space
				startTexFig \space }
		\if@angle
			\special {ps:: \@p@sangle \space rotate \space} 
		\fi
		\if@clip{
			\if@verbose{
				\typeout{(clip)}
			}\fi
			\special{ps:: doclip \space }
		}\fi
		\if@prologfile
		    \special{ps: plotfile \@prologfileval \space } \fi
		\special{ps: plotfile \@p@sfile \space }
		\if@postlogfile
		    \special{ps: plotfile \@postlogfileval \space } \fi
		\special{ps::[end] endTexFig \space }
		\vbox to \@p@srheight true sp{
			\hbox to \@p@srwidth true sp{
				\hss
			}
		\vss
		}
	}\else{
		\if@draftbox{		
			\hbox{\fbox{\vbox to \@p@srheight true sp{
			\vss
			\hbox to \@p@srwidth true sp{ \hss \@p@sfile \hss }
			\vss
			}}}
		}\else{
			\vbox to \@p@srheight true sp{
			\vss
			\hbox to \@p@srwidth true sp{\hss}
			\vss
			}
		}\fi

	}\fi
}}
\def\psglobal{\typeout{psfig: PSGLOBAL is OBSOLETE; use psprint -m instead}}
\psfigRestoreAt



 
\numerazionedoppia
\autobibliografia
\def\cost{\hbox{\rm const.}} 
\def\mes{\hbox{\rm{ mes}}\,}
\def\IM{\hbox{\rm{Im}}\,}
\def\ID{{\bf 1}}
\def\RE{\hbox{\rm{Re}}\,}
\def\Id{\hbox{\rm{Id}}\,}
\def\ddx{{{\rm d}\over{\rm d} x}}
\def\ddtx{{{\rm d}\over{\rm d}\hat x}}
\def\ddz{{{\rm d}\over{\rm d} z}}
\def\thi{{\vartheta_\infty}} 
\def\eps{\varepsilon}
\def\phi{\varphi} 

\centerline{\bf RATIONAL SOLUTIONS OF THE PAINLEVE'--VI EQUATION}
\vskip 0.3 cm
\centerline{\bf M. Mazzocco}
\vskip 0.2 cm
\centerline{\it Mathematical Institute}

\centerline{\it24-29 St Giles, Oxford, UK.}

\vskip 0.5 cm

\noindent{\bf Abstract.}\quad 
In this paper, we classify all values of the parameters $\alpha$,
$\beta$, $\gamma$ and $\delta$ of the Painlev\'e VI equation such that 
there are rational solutions. We give a formula for them up to 
the birational canonical transformations and the symmetries of the 
Painlev\'e VI equation.

\vskip 0.5 cm
\semiautosez{1}{1. Introduction.}

In this paper, we study the general Painlev\'e sixth equation
$$
\eqalign{
y_{xx}=&{1\over2}\left({1\over y}+{1\over y-1}+{1\over y-x}\right) y_x^2 -
\left({1\over x}+{1\over x-1}+{1\over y-x}\right)y_x+\cr
&+{y(y-1)(y-x)\over x^2(x-1)^2}\left[\alpha+\beta {x\over y^2}+
\gamma{x-1\over(y-1)^2}+\delta {x(x-1)\over(y-x)^2}\right],\cr}
\eqno{PVI}
$$
where $x\in\complessi$ and $\alpha,\beta,\gamma,\delta$ are arbitrary
complex parameters.
The general solution $y(x;c_1,c_2)$ of the PVI equation 
satisfies the following two important properties (see [Pain], [Gam]):
\vskip 0.2 cm
\item{1)} {\it Painlev\'e property:}\/ the solutions $y(x;c_1,c_2)$ may have 
complicated singularities (i.e. branch points, essential singularities etc.) 
only at the {\it critical points}\/ $0,1,\infty$ of the equation  
(the so-called {\it fixed singularities}\/). All the other singularities, the
position of which depend on the integration constants (the so-called 
{\it movable singularities}\/), are poles. 

\item{2)} For generic values of the integration constants $c_1,c_2$ and of 
the parameters $\alpha,\beta,\gamma,\delta$, the solution $y(x;c_1,c_2)$ can 
not be expressed via {\it known}\/ functions.
\vskip 0.2 cm

The latter property needs to be stated more precisely. Following [Um1],
[Um2], we define known or {\it classical functions}\/ to be functions 
that can be obtained from the field of rational functions
$\complessi(x)$, by a finite iteration of the following operations:
\item{i)} derivation,
\item{ii)} quadrature,
\item{iii)} arithmetic operations $+,-,\times,\div$, 
\item{iv)} solution of a homogeneous linear ordinary differential
equations with classical coefficients,
\item{v)} substitution into an Abelian function,
\item{vi)} solution of algebraic differential
equations of first order with classical coefficients.
\vskip 0.2 cm

According to this definition, Watanabe (see [Wat]) proved that for
generic values of the integration constants $c_1,c_2$ and of 
the parameters $\alpha,\beta,\gamma,\delta$, the solutions
$y(x;c_1,c_2)$ of PVI are non-classical and classified all the
one-parameter families of classical solutions of PVI. Loosely speaking,
Watanabe proves that all one-parameter families of classical solutions
of PVI are realized for values of the parameters
$\alpha,\beta,\gamma,\delta$ lying on the walls of the Weyl chamber
of the group $\tilde W$ of the birational canonical 
transformations.\footnote{${}^{1}$}{Recall that $\tilde W$ is
isomorphic to $W_a(D_4)$ the affine extension of the Weyl group of $D_4$,
(see [Ok]).} Such one-parameter families of classical solutions, 
already found by Okamoto, shrink down, by the action of the group 
$\tilde W$, to the following list\footnote{${}^{2}$}{The group $
\tilde W$ acts on $y(x)$ and on its conjugate momentum $p(x)$. 
In the list i)$\dots$,iv) the conjugate momentum $p(x)$ is given by a 
one-parameter family.}
\item{i)} $y(x)\equiv \infty$, for $\alpha=0$,
\item{ii)} $y(x)\equiv 0$, for $\beta=0$,
\item{iii)} $y(x)\equiv 1$, for $\gamma=0$,
\item{iv)} $y(x)\equiv x$, for $\delta={1\over2}$,
\item{v)}  {\it Riccati solutions,}\/ 
$$
\ddx y= -{y(y-1)(y-x)\over x(x-1)} \left({\vartheta_1\over y}+
{\vartheta_2-1\over y-x}+{\vartheta_3\over y-1}\right),\autoeqno{wat}
$$
for $\thi= \vartheta_1+\vartheta_2+\vartheta_3$, where 
$\thi,\vartheta_1,\vartheta_2,\vartheta_3$ are defined by
$$
\alpha={(\thi-1)^2\over2},\quad\beta=-{\vartheta_1^2\over2},
\quad\gamma={\vartheta_3^2\over2},\quad\delta={1-\vartheta_2^2\over2}.
\autoeqno{wat1}
$$
Solutions (i), (ii), (iii), (iv) are called degenerate.

The theory of the rational and classical solutions of the Painlev\'e
sixth equation has been extensively studied in [Ai], [AMM], 
[Gr], [GL], [Luk], [Um], [Wat]. In this paper we classify all rational 
solutions of PVI. We prove the following

\proclaim Main Theorem. All rational non-degenerate solutions of PVI 
belong to the 
intersection of two or more one-parameter families of classical
solutions, i.e. they occur for
$$
\thi+\sum_{k=1}^3\eps_k\vartheta_k\in2\interi,\quad \eps_k=\pm1,
\quad\hbox{and}\quad
\vartheta_k\in\interi\,\hbox{ for at least one}\quad k=1,2,3,\infty.
$$
All rational non-degenerate solutions of the PVI equation are equivalent via 
birational canonical transformations and up to 
symmetries\footnote{${}^{2}$}{\rm The symmetries of the
Painlev\'e VI equation are compositions of the following transformations
i) $x\to 1-x$, $y\to 1-y$, 
$\vartheta_1\leftrightarrow\vartheta_3$, ii) $x\to{1\over x}$,
$y\to{1\over y}$, $\thi\leftrightarrow\vartheta_1+1$, iii)  
$x\to{1\over 1-x}$, $y\to{q-x\over x-1}$, 
$\vartheta_1\leftrightarrow\vartheta_2$.}
to the following solutions
$$
y(x)={(x\over 
(1+\vartheta_2+x+x\vartheta_3)}
,\quad\hbox{for}\quad 
\thi=-\sum_{k=1}^3\vartheta_k\quad\hbox{and}\quad \vartheta_1=1,
$$
$$
y(x)={(\vartheta_2+\vartheta_3 x)^2-\vartheta_2-\vartheta_3 x^2\over
(\vartheta_2+\vartheta_3-1)(\vartheta_2+\vartheta_3 x)},
\quad\hbox{for}\quad 
\thi=-\sum_{k=1}^3\vartheta_k\quad\hbox{and}\quad \vartheta_1=-2.
$$

Our method to prove this result does not use Umemura's theory, but the
isomonodromy deformation method (see [Fuchs], [Sch], [JMU], [ItN], [FlN]). 
The Painlev\'e VI is represented as the equation of isomonodromy deformation 
of the two-dimensional auxiliary Fuchsian system
$$
{{\rm d}\over{\rm d}\lambda} \Phi=\left({{\cal A}_1\over \lambda-u_1} + 
{{\cal A}_2\over \lambda-u_2} + {{\cal A}_3\over \lambda-u_3}\right)\Phi,
\autoeqno{in1}
$$
${\cal A}_j$ being $2\times 2$ matrices independent on $\lambda$, and 
$u_1,u_2,u_3$ being pairwise distinct complex numbers. 
The matrices ${\cal A}_j$ satisfy the following conditions:
$$
{\rm eigen}\left({\cal A}_j\right)=\pm{\vartheta_j\over2}
\quad\hbox{and}\quad
-{\cal A}_1-{\cal A}_2-{\cal A}_3={\cal A}_\infty:=
{1\over2}\pmatrix
{\thi & 0\cr 0 & -\thi\cr},
$$
$\vartheta_j$, $j=1,2,3,\infty$ being related to the
parameters $\alpha,\beta,\gamma,\delta$ of PVI as in \eqrefp{wat1}.

The entries of the matrices $A_i$ are complicated expressions of $x,y,y_x$ 
and of some quadrature $\int R(x,y){\rm d}x$, $x$ being the cross
ratio of the poles $x={u_2-u_1\over u_3-u_1}$. The monodromy matrices 
${\cal M}_1$, ${\cal M}_2$ and ${\cal M}_3$ of \eqrefp{in1} remain 
constant if and only if $y=y(x)$ satisfies PVI. To each branch of a 
solution of PVI corresponds a triple ${\cal M}_1,{\cal M}_2,{\cal M}_3$
of monodromy matrices, which is unique up to
$$
\left({\cal M}_1,{\cal M}_2,{\cal M}_3\right)
\to L_\infty^{-1}\left({\cal M}_1,{\cal M}_2,{\cal M}_3\right)
L_\infty,
$$
where $L_\infty$ is any constant invertible matrix such that 
$[L_\infty,{\cal M}_\infty]=0$ (see Section 2).

Following the same strategy as in [DM], we describe the procedure of 
analytic continuation of a branch of a solution of PVI by the action 
of the pure braid group on its monodromy matrices. Since rational 
solutions have only one branch, we look for fixed points of this action.
We show that a necessary condition to have a rational (non-degenerate) 
solution is that the corresponding monodromy group is abelian (see Section 3). 

Abelian $2\times2$ monodromy groups are reducible. In Section 4, we
classify all solutions of PVI having a reducible monodromy group (such 
solutions where found in [Hit] as a reduction of Nahm's equations for 
a diffeomorphic group). Then we classify all rational solutions among them. 

\vskip 0.5 cm
\noindent{\bf Acknowledgements.} The idea of classifying all rational
solutions of PVI came out of a conversation with H. Umemura.
I am indebted to B. Dubrovin who introduced me to the theory of Painlev\'e 
equations and gave me lots of suggestions. I am grateful to
N. Hitchin, who constantly addressed my work and A.C.C. Coolen 
for kindly hosting me at Kings College London where this 
paper was started. The author was supported by an EPSRC research 
assistanship. We thank P. Clarkson and P. Boalch for pointing out some of
the typos which have been fixed in this version.

\vskip 0.4 cm
\semiautosez{2}{\bf 2. The Painlev\'e VI equation as the isomonodromic
deformation equation of a $2\times2$ Fuchsian system.}

Consider the following Fuchsian system with four pairwise distinct  
regular singularities at $u_1,u_2,u_3,\infty$:
$$
{{\rm d}\over{\rm d}\lambda} \Phi=\left({{\cal A}_1\over \lambda-u_1} + 
{{\cal A}_2\over \lambda-u_2} + {{\cal A}_3\over \lambda-u_3}\right)
\Phi,\qquad \lambda\in\overline\complessi\backslash\{u_1,u_2,u_3,\infty\}
\autoeqno{N1}
$$
${\cal A}_j$ being $2\times 2$ matrices independent of $\lambda$. Assume that 
the matrices ${\cal A}_j$ satisfy the following conditions:
$$
{\rm eigen}\left({\cal A}_j\right)=\pm{\vartheta_j\over2}
\quad\hbox{and}\quad
-{\cal A}_1-{\cal A}_2-{\cal A}_3={\cal A}_\infty,\autoeqno{N1.3}
$$
for some constants $\vartheta_j$, $j=1,2,3$ and
$$
{\cal A}_\infty:={1\over2}\pmatrix
{\thi & 0\cr 0 & -\thi\cr},\qquad\hbox{
for some constant}\, \thi\neq 0.
\autoeqno{gigio}
$$
The solution $\Phi(\lambda)$ of the system \eqrefp{N1} is a multi-valued 
analytic function on $\complessi\backslash\{u_1,u_2,u_3\}$, and its
multivaluedness is described by the {\it monodromy matrices.}\/ 
To define them, we fix a basis $\gamma_1,\gamma_2,\gamma_3$ of loops
in $\pi_1\left(\overline\complessi\backslash\{u_1,u_2,u_3,\infty\},
\infty\right)$ as in figure 1, and a fundamental matrix for the 
system \eqrefp{N1}.

\midinsert\centerline{\psfig{file=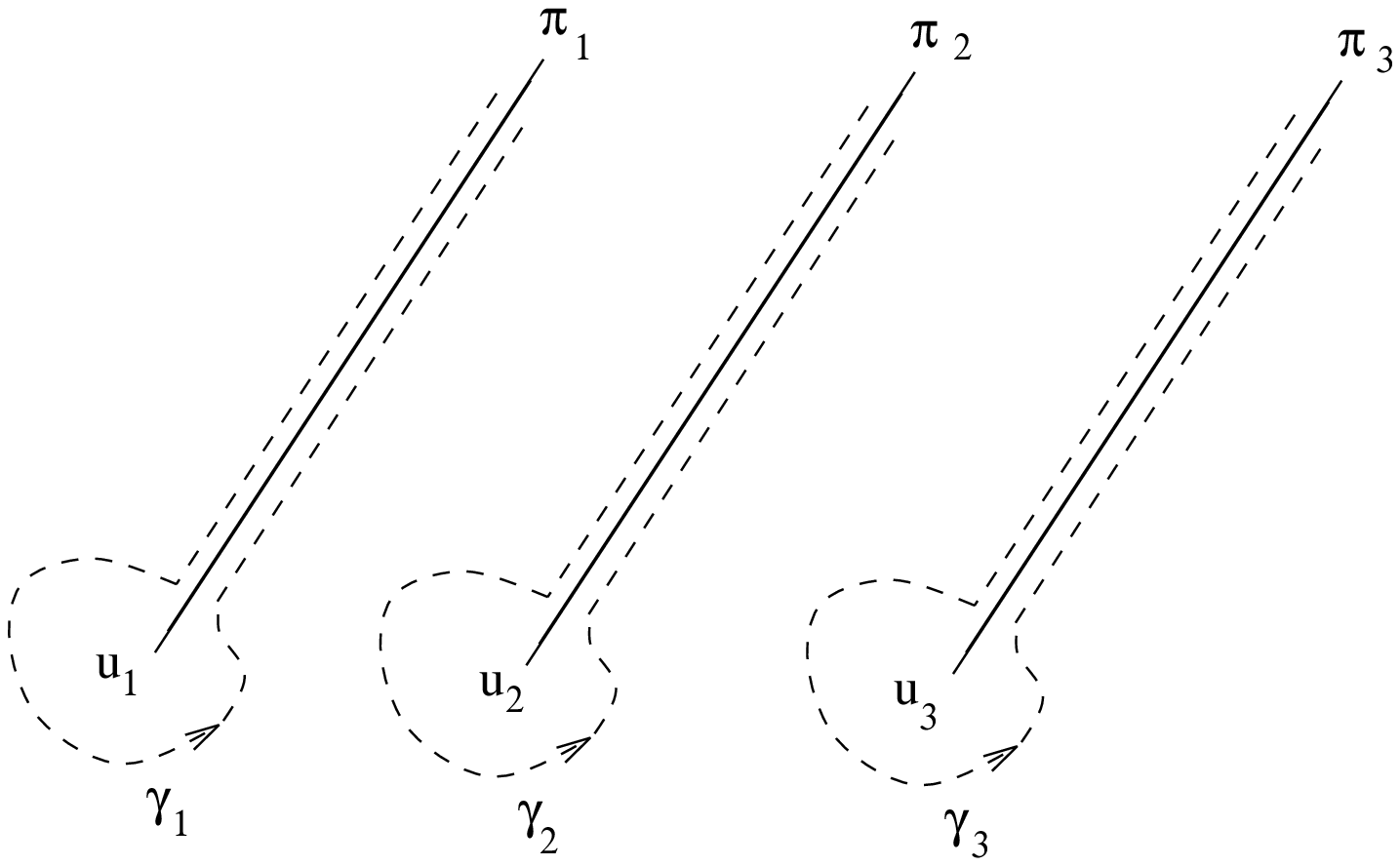,height=3cm}}
\vskip 0.6 cm
\noindent {{\bf Fig. 1.} The branch-cuts $\pi_j$ between the ordered
singularities $u_j$ and the oriented loops $\gamma_j$ in the $\lambda$-plane.}
\endinsert

\proclaim Proposition 2.1. There exists a fundamental matrix of the system 
\eqrefp{N1} of the form
$$
\Phi_\infty(\lambda)=\left(\ID +{\cal O}({1\over \lambda})\right) 
\lambda^{-{\cal A}_\infty}\lambda^{-{\cal R}_\infty},\quad\hbox{as}\quad 
\lambda\rightarrow\infty,
\autoeqno{M3}
$$
where $\lambda^{-{\cal R}_\infty}:=e^{-{\cal R}_\infty\log\lambda}$, with 
the choice of the principal branch of the logarithm with the
branch-cut along the common direction $\eta$ of the cuts
$\pi_1,\pi_2,\pi_3$ and the matrix entries of ${{\cal R}_\infty}$
given by ${{\cal R}_\infty}_{11}={{\cal R}_\infty}_{22}=0$ and
$$
\eqalign{
&\hbox{for}\,\thi\not\in\interi,\,\hbox{and for }\,\thi =0, 
\quad {{\cal R}_\infty}_{12}={{\cal R}_\infty}_{21}=0,\cr
&\hbox{for}\,\thi=n\in\interi_+,\quad 
{{\cal R}_\infty}_{12}={{\cal R}_\infty}_{12}({\cal A}_{1,2,3},u_{1,2,3}),
\quad {{\cal R}_\infty}_{21}=0,\cr
&\hbox{for}\,\thi=-n,\,n\in\interi_+,\quad
{{\cal R}_\infty}_{21}={{\cal R}_\infty}_{21}({\cal A}_{1,2,3},u_{1,2,3}),
\quad {{\cal R}_\infty}_{12}=0,\cr}\autoeqno{M4}
$$
where the functions ${{\cal R}_\infty}_{12,21}({\cal A}_{1,2,3},u_{1,2,3})$
are uniquely determined by $({\cal A}_{1,2,3},u_{1,2,3})$.
Such a fundamental matrix $\Phi_\infty(\lambda)$ is uniquely
determined up to 
$$
\Phi_\infty(\lambda)\to \Phi_\infty(\lambda)L_\infty,\autoeqno{amb}
$$
where $L_\infty$ is any constant invertible matrix such that
$$
\lambda^{-{\cal A}_\infty}\lambda^{-{\cal R}_\infty}L_\infty
\lambda^{{\cal A}_\infty}\lambda^{{\cal R}_\infty}=
\ID + \sum_{k=1}^N {L_\infty^{(k)}\over\lambda^k},
\autoeqno{amb1}
$$
for some $L_\infty^{(1)},\cdots,L_\infty^{(N)}$ constant matrices.

\noindent The proof can be found in [Dub]. 
\vskip 0.2 cm 

The fundamental matrix $\Phi_\infty$ can be analytically 
continued to an analytic function on the universal covering of 
$\overline\complessi\backslash\{u_1,u_2,u_3,\infty\}$. For any element 
$\gamma\in\pi_1
\left(\overline\complessi\backslash\{u_1,u_2,u_3,\infty\},\infty\right)$
denote the result of the analytic continuation of $\Phi_\infty(\lambda)$ 
along the loop $\gamma$ by $\gamma[\Phi_\infty(\lambda)]$. Since 
$\gamma[\Phi_\infty(\lambda)]$ and $\Phi_\infty(\lambda)$ are two fundamental 
matrices in the neighbourhood of infinity, they are related by the 
following relation:
$$
\gamma[\Phi_\infty(\lambda)]= \Phi_\infty(\lambda) {\cal M}_\gamma
$$
for some constant invertible $2\times 2$ matrix ${\cal M}_\gamma$
depending only 
on the homotopy class of $\gamma$. Particularly, the matrix 
${\cal M}_\infty:={\cal M}_{\gamma_\infty}$, $\gamma_\infty$ being a
simple loop around 
infinity in the clock-wise direction, is given by:
$$
{\cal M}_\infty=\exp(2\pi i {\cal A}_\infty)
\exp(2\pi i{{\cal R}_\infty}).
\autoeqno{N9}
$$
The resulting {\it monodromy representation}\/ is an anti-homomorphism:
$$
\matrix{
\pi_1\left(\overline\complessi\backslash\{u_1,u_2,u_3,\infty\},\infty\right)
&\rightarrow & SL_2(\complessi)\cr
\gamma&\mapsto & {\cal M}_\gamma\cr}\autoeqno{N2.2}
$$
$$
{\cal M}_{\gamma\tilde\gamma}= {\cal M}_{\tilde\gamma}  
{\cal M}_{\gamma}. \autoeqno{N2}
$$ 
The images ${\cal M}_j:={\cal M}_{\gamma_j}$ of the generators 
$\gamma_j$, $j=1,2,3$ of 
the fundamental group, are called {\it the monodromy matrices}\/ of 
the Fuchsian system \eqrefp{N1}. They generate the 
{\it monodromy group of the system,}\/ i.e. the image of the representation 
\eqrefp{N2.2}. Since the loop $(\gamma_1 \gamma_2 \gamma_3)^{-1}$ is
homotopic to $\gamma_\infty$, the following relation between the
generators holds:
$$
{\cal M}_\infty {\cal M}_3 {\cal M}_2 {\cal M}_1=\ID.\autoeqno{N6}
$$
Observe that if we fix another fundamental matrix 
$\Phi_\infty'=\Phi_\infty L_\infty$ in the equivalence class defined 
by \eqrefp{amb}, the monodromy matrices ${\cal M}_\gamma'$ with
respect to the new fundamental matrix $\Phi_\infty'$ are related to
the old ones by
$$
{\cal M}_j'= L_\infty^{-1}{\cal M}_jL_\infty,\quad
j=1,2,3.
$$
Thus the monodromy matrices ${\cal M}_3$, ${\cal M}_2$, ${\cal M}_1$ are
uniquely defined up to the ambiguity
$$
({\cal M}_1, {\cal M}_2, {\cal M}_3)\sim
(L_\infty^{-1}{\cal M}_1L_\infty,L_\infty^{-1} {\cal M}_2L_\infty, 
L_\infty^{-1}{\cal M}_3L_\infty),\autoeqno{amb2}
$$
where $L_\infty$ is given by \eqrefp{amb1}. Observe that ${\cal M}_\infty$ 
is invariant w.r.t. \eqrefp{amb2}.

I recall the definition of the {\it connection matrices.}\/ Near the poles 
$u_j$, the fundamental matrices $\Phi_j(\lambda)$ of the 
system \eqrefp{N1}, are given by the following

\proclaim Proposition 2.2. There exists a fundamental matrix of the system 
\eqrefp{N1} of the form
$$
\Phi_j(\lambda)=G_j\left(\ID +{\cal O}(\lambda-u_j)\right) 
(\lambda-u_j)^{J_j}(\lambda-u_j)^{{\cal R}_j},\quad\hbox{as}\quad 
\lambda\rightarrow u_j,
\autoeqno{N6.1}
$$
where
$$
\eqalign{
&\hbox{for }\vartheta_j\neq0,\qquad
J_j={1\over2}\pmatrix{\vartheta_j&0\cr 0 & -\vartheta_j\cr},\cr
&\hbox{for } \vartheta_j=0, \qquad
J_j=J\equiv\pmatrix{0&1\cr 0 & 0\cr}.\cr}
$$
The invertible matrix $G_j$ is defined by ${\cal A}_j=G_j J_j G_j^{-1}$, 
the diagonal elements of the matrix ${\cal R}_j$ are zero and the
off-diagonal ones are defined as follows,
$$
\eqalign{
&\hbox{for}\,\vartheta_j\not\in\interi,\,\hbox{and for }\,\vartheta_j=0 \quad 
{{\cal R}_j}_{12}={{\cal R}_j}_{21}=0,\cr
&\hbox{for}\,\vartheta_j=n\in\interi_+,\quad 
{{\cal R}_j}_{12}={{\cal R}_j}_{12}({\cal A}_{1,2,3},u_{1,2,3}),
\quad {{\cal R}_j}_{21}=0,\cr
&\hbox{for}\,\vartheta_j=- n,\, n\in\interi_+,\quad
{{\cal R}_j}_{21}={{\cal R}_j}_{21}({\cal A}_{1,2,3},u_{1,2,3}),
\quad{{\cal R}_j}_{12}=0,\cr}\autoeqno{M5}
$$
where the functions ${{\cal R}_j}_{21,12}({\cal A}_{1,2,3},u_{1,2,3})$ 
are uniquely determined by ${\cal A}_{1,2,3}$ and $u_{1,2,3}$.
The choice
of the branch of $\log(z-u_j)$ is along $\eta$ as
above. The fundamental matrix $\Phi_j(\lambda)$ is uniquely
determined up to the ambiguity:
$$
\Phi_j(\lambda)\mapsto \Phi_j(\lambda)L_j\autoeqno{amb45}
$$
where $L_j$ is any constant invertible matrix such that 
$$
(\lambda-u_j)^{J_j}(\lambda-u_j)^{{\cal R}_j}L_j
(\lambda-u_j)^{-J_j}(\lambda-u_j)^{-{\cal R}_j}=
\sum_{k=0}^N {L_j^{(k)}(\lambda-u_j)^k},\autoeqno{amb5}
$$
for $L_j^{(0)}={G}_j$ and for some $L_j^{(1)},\cdots,L_j^{(N)}$ constant
matrices.

\noindent The proof can be found in [Dub].
\vskip 0.3 cm

Continuing the solution $\Phi_\infty(\lambda)$ to a neighbourhood of $u_j$, 
along, say, the right-hand-side of the branch-cut $\pi_j$, one obtains another 
fundamental matrix around $u_j$, that must be related to $\Phi_j(\lambda)$ by:
$$
\Phi_\infty(\lambda)=\Phi_j(\lambda){\cal C}_j,\autoeqno{N3}
$$
for some invertible matrix ${\cal C}_j$. The matrices 
${\cal C}_1,{\cal C}_2, {\cal C}_3$  are called 
{\it connection matrices,}\/ and they are defined by \eqrefp{N3} up to
the ambiguity ${\cal C}_j\to{\cal C}_jL_\infty$ due to \eqrefp{amb}.
The connection matrices are related to the monodromy matrices as 
follows:
$$
{\cal M}_j={\cal C}_j^{-1} \exp\left(2\pi i J_j\right)
\exp\left( {\cal R}_j\right) 
{\cal C}_j,\qquad j=1,2,3.
\autoeqno{N4}
$$
Thanks to the above relation it follows that
$$
{\rm eigen}({\cal M}_j)=\exp(\pm\pi i\vartheta_j).\autoeqno{N5}
$$

\proclaim Definition 2.3. The {\it Monodromy data}\/ of the Fuchsian 
system \eqrefp{N1} are 
$$
\left\{({\cal M}_1,{\cal M}_2,{\cal M}_3)
\slash_\sim,\,
{\cal R}_1,{\cal R}_2,{\cal R}_3\right\},
$$
where $\sim$ is the equivalence relation defined by \eqrefp{amb2}.

\noindent{\bf Remark 2.4.} For non-integer $\vartheta_j$, the correspondent
${\cal R}_j$ matrix is zero by definition and we drop it from the set of the 
monodromy data.
\vskip 0.2 cm

The theory of the deformations the poles of the Fuchsian system keeping the 
monodromy fixed is described by the following result:

\proclaim Theorem 2.5. Let $\left\{({\cal M}_1,{\cal M}_2,{\cal M}_3)
\slash_\sim, \, {\cal R}_1,{\cal R}_2,{\cal R}_3\right\}$, be
monodromy data of the Fuchsian system:
$$
{{\rm d}\over{\rm d}\lambda} \Phi^0=
\left({{\cal A}^0_1\over \lambda-u^0_1}+{{\cal A}^0_2\over \lambda-u^0_2}+
{{\cal A}^0_3\over \lambda-u^0_3}\right)\Phi^0,
\autoeqno{N7}
$$
of the above form \eqrefp{N1.3}, with pairwise distinct poles $u_j^0$, 
and with respect to some basis $\gamma_1,\gamma_2,\gamma_3$ of the loops in 
$\pi_1\left(\overline\complessi\backslash\{u^0_1,u^0_2,u^0_3,\infty\},
\infty\right)$. If ${\cal M}_k\neq\pm\ID$ for all $k=1,2,3,\infty$, 
there exists a neighbourhood 
$U\subset\complessi^3$ of the point $u^0=(u^0_1,u^0_2,u^0_3)$ such
that, for any $u= (u_1,u_2,u_3)\in U$, there exists a unique triple 
${\cal A}_1(u)$, $ {\cal A}_2(u)$, $ {\cal A}_3(u)$ of analytic matrix
valued functions such that:
$$
{\cal A}_j(u^0)= {\cal A}_j^0,\quad i=1,2,3,
$$
and the monodromy matrices of the Fuchsian system
$$
{{\rm d}\over{\rm d}\lambda} \Phi= A(\lambda;u) \Phi=
\left({ {\cal A}_1(u)\over \lambda-u_1}+{ {\cal A}_2(u)\over \lambda-u_2}+
{ {\cal A}_3(u)\over \lambda-u_3}\right)\Phi, \autoeqno{N8}
$$
with respect to the same basis\footnote{${}^{1}$}{Observe that the 
basis $\gamma_1,\gamma_2,\gamma_3$ of 
$\pi_1\left(\overline\complessi\backslash\{u_1,u_2,u_3,\infty\},
\infty\right)$ varies continuously with small variations of $u_1,u_2,u_3$. 
This new basis is homotopic to the initial one, so one can identify them.}
$\gamma_1,\gamma_2,\gamma_3$ of the loops, can be chosen to 
coincide with the given ${\cal M}_1$, ${\cal M}_2$, ${\cal M}_3$.
The matrices ${\cal A}_j(u)$ are the solutions of the Cauchy problem with 
the initial data $ {\cal A}_j^0$ for the following Schlesinger equations:
$$
{\partial\over\partial u_j} {\cal A}_i= 
{[ {\cal A}_i, {\cal A}_j]\over u_i-u_j},\quad
{\partial\over\partial u_i} {\cal A}_i= 
-\sum_{j\neq i}{[ {\cal A}_i, {\cal A}_j]\over u_i-u_j}. 
\autoeqno{N10}
$$
The solution $\Phi_\infty^0(\lambda)$ of \eqrefp{N7} of the form 
\eqrefp{M3} can be uniquely continued, for $\lambda\neq u_i$ $i=1,2,3$, 
to an analytic function $\Phi_\infty(\lambda,u),\quad u\in U$, such that
$$
\Phi_\infty(\lambda,u^0)=\Phi_\infty^0(\lambda).
$$
This continuation is the local solution of the Cauchy problem with 
the initial data $\Phi_\infty^0$ for the following system that is 
compatible with the system \eqrefp{N8}:
$$
{\partial\over\partial u_i} \Phi = -{ {\cal A}_i(u)\over \lambda-u_i} \Phi.
$$
Moreover the functions $ {\cal A}_i(u)$ and $\Phi_\infty(\lambda,u)$ can be 
continued analytically to global meromorphic functions on the universal 
coverings of
$$
\complessi^3\backslash\{diags\}:=
\left\{(u_1,u_2,u_3)\in\complessi^3\,|\,u_i\neq u_j\,\hbox{for}\, 
i\neq j\right\},
$$
and
$$
\left\{(\lambda,u_1,u_2,u_3)\in\complessi^4\,|\,u_i\neq u_j\,\hbox{for}\, 
i\neq j\,\hbox{and}\,\lambda\neq u_i,\, i=1,2,3\right\},
$$
respectively.

The proof of this theorem can be found, for example, in [Mal], [Miwa], 
[Sib].

\vskip 0.2 cm
\noindent{\bf Remark 2.6.} Observe that the isomonodromic deformations 
equations preserve the connection matrices ${\cal C}_i$ too.
\vskip 0.2 cm

\noindent{\bf Remark 2.7.} When ${\cal M}_k=\pm\ID$ for some 
$k=1,2,3\infty$, the existence statements of Theorem 2.5 are still
valid, while the uniqueness ones are lost. 
\vskip 0.2 cm

Let me now explain, following [JMU], how to rewrite the Schlesinger equations 
\eqrefp{N10} in terms of the PVI equation. We can assume $\thi\neq 0$
without loss of generality (see Remark 2.9 below).
Schlesinger equations \eqrefp{N10} with fixed ${\cal A}_\infty$ are 
invariant with respect to the gauge transformations of the form:
$$
{\cal A}_i\mapsto D^{-1} {\cal A}_i D,\quad i=1,2,3,\quad\hbox{for any $D$ 
diagonal matrix}.\autoeqno{dig}
$$ 
We introduce two coordinates $(p,q)$ on the quotient of the space of  
matrices satisfying \eqrefp{N10} with respect to the equivalence relation
\eqrefp{dig} and a coordinate $k$ that contains the above gauge 
freedom:
$$
[{\cal A}(q;u_1,u_2,u_3)]_{12}=0, \qquad
p=\sum_{k=1}^3{{\cal A}_{k_{11}}+{\vartheta_k\over2}\over q-u_k},\qquad
k={2 P(\lambda)[{\cal A}(\lambda;u_1,u_2,u_3)]_{12}
\over\thi(q-\lambda)},
$$
where ${\cal A}(z;u_1,u_2,u_3)$ is given in \eqrefp{N8} and 
$P(\lambda)=(\lambda-u_1)(\lambda-u_2)(\lambda-u_3)$. The matrices 
${\cal A}_i$ are uniquely determined by the coordinates $(p,q)$, and
$k$ and expressed rationally in terms of them:
$$
\eqalign{
{{\cal A}_i}_{11}&
={1\over\thi P'(u_i)}\bigg\{P(q)(q-u_i)p^2+P(q)(q-u_i)p
\left({\thi\over q-u_i}-\sum_{k=1}^3{\vartheta_k\over q-u_k}\right)+\cr
&+(q-u_i)\left[{\thi^2\over4}\left(q+2 u_i-\sum_{k=1}^3 u_k\right)+
\sum_{k=1}^3{\vartheta_k^2\over 4}
\left(q+2 u_k-\sum_{j=1}^3 u_j\right)\right]+\cr
&+{q-u_i\over2} \left(\vartheta_1 \vartheta_2 (q-u_3)+
\vartheta_1\vartheta_3 (q-u_2)+
\vartheta_2\vartheta_3(q-u_1)\right)-
{\thi\over2}\sum_{k=1}^3{\vartheta_k\over q-u_k}
\bigg\}\cr
{{\cal A}_i}_{12}&=-\thi k {q-u_i\over 2 P'(u_i)},\cr
{{\cal A}_i}_{21}&= {1\over {{\cal A}_i}_{12}}
\left({\vartheta_i^2\over 4}-{{\cal A}_i}_{11}^2 \right),\cr
{{\cal A}_i}_{22}&=-{{\cal A}_i}_{11}
\cr}\autoeqno{N12.5}
$$
for $i=1,2,3$, where $P'(z)={{\rm d}P\over{\rm d}z}$. 
The Schlesinger equations \eqrefp{N10} in these variables are:
$$
\left\{
\eqalign{
{\partial q\over\partial u_i} &=
{P(q)\over P'(u_i)}\left[2 p + {1\over q-u_i}-
\sum_{k=1}^3{\vartheta_k\over q-u_k}\right]\cr
{\partial p\over\partial u_i} &=-\bigg\{
P'(q) p^2 +
\big[2q+u_i-\sum_ju_j-\sum_{k=1}^3
\vartheta_k(2q+u_k-\sum_ju_j)\big]p+\cr
&+{1\over4}(\sum_{k=1}^3\vartheta_k-\thi)
(\sum_{k=1}^3\vartheta_k+\thi-2)\bigg\}
{1\over P'(u_i)},
\cr}\right.
\autoeqno{N13}
$$
and
$$
{\partial\log(k)\over\partial u_i}=(\thi-1){q-u_i\over P'(u_i)}.
\autoeqno{N13.5}
$$
for $i=1,2,3$. The system of the {\it reduced Schlesinger equations}\/ 
\eqrefp{N13} is invariant under the transformations of the form
$$
u_i\mapsto a u_i + b,\qquad
q\mapsto a q + b,\qquad
p\mapsto {p\over a}, \qquad\forall a,b\in\complessi,\quad a\neq0.
$$
Introducing the following new invariant variables:
$$
x={u_2-u_1\over u_3-u_1},\quad
y={q-u_1\over u_3-u_1};\autoeqno{N13.6}
$$
the system \eqrefp{N13}, expressed in the these new variables, gives 
the PVI equation for $y(x)$ with parameters 
$$
\alpha={(\thi-1)^2\over2},\quad\beta=-{\vartheta_1^2\over2},
\quad\gamma={\vartheta_3^2\over2},\quad\delta={1-\vartheta_2^2\over2}.
\autoeqno{8}
$$

\vskip 0.2 cm
\noindent{\bf Remark 2.8.}\quad Observe that permutations of the poles $u_i$ 
and of the values $\vartheta_i$, $i=1,2,3,\infty$
induce transformations of $(y,x)$ of the type $x\to1-x$ and $y\to1-y$, 
$x\to{1\over x}$ and $y\to{1\over y}$, $x\to{1\over 1-x}$ and
$y\to{y-x\over 1-x}$ and their compositions. These
transformations are the {\it symmetries}\/ of the Painlev\`e VI equation.
 
\vskip 0.2 cm
\noindent{\bf Remark 2.9.}\quad It is clear from \eqrefp{8} that
changes of the signs of the parameters $\vartheta_k$, $k=1,2,3$ and
transformations on $\thi$ of type $\thi\to2-\thi$ give rise to the same
PVI equation. 

\vskip 0.2 cm
\noindent We summarise the results of this section in the following:

\proclaim Theorem 2.10. Branches $y(x)$ of solutions of the PVI
equation with parameters $\alpha$, $\beta$, $\gamma$, $\delta$,
considered up to symmetries are in one to one correspondence with 
local solutions of the Schlesinger equations \eqrefp{N10} with 
parameters $\vartheta_1,\vartheta_2,\vartheta_3,\thi$ given by
\eqrefp{8} and ${\cal A}_\infty$ given in \eqrefp{gigio},
considered up to diagonal conjugation \eqrefp{dig} and permutation. 
This one-to-one correspondence is realized by the formulae 
\eqrefp{N12.5}, \eqrefp{N13.6}. For each branch of a solution of the
PVI equation, there exist a unique set of monodromy data
$\left\{({\cal M}_1,{\cal M}_2,{\cal M}_3)\slash_\sim,
{\cal R}_1,{\cal R}_2,{\cal R}_3\right\}$.
Vice-versa, let
$\left\{({\cal M}_1,{\cal M}_2,{\cal M}_3)\slash_\sim,
{\cal R}_1,{\cal R}_2,{\cal R}_3\right\}$ be a set of monodromy data such that
$$
{\rm eigen}\left({\cal M}_j\right)=\exp(\pm\pi i\vartheta_j),
\qquad
({\cal M}_3 {\cal M}_2 {\cal M}_1)^{-1}={\cal M}_\infty,
\quad {\rm eigen}\left({\cal M}_\infty\right)=\exp(\pm\pi i\thi),
$$
$$
{\cal M}_k\neq\pm\ID\qquad \forall\quad k=\,1,2,3,\infty,
$$
with some numbers $\vartheta_1,\vartheta_2,\vartheta_3,\thi$,
and ${\cal R}_1,{\cal R}_2,{\cal R}_3$ satisfying (\eqref{M5}).
If there exists
a branch of a solution of the PVI equation with parameters \eqrefp{8}
such that the Fuchsian system of the form \eqrefp{N1} given by 
\eqrefp{N12.5} has the prescribed monodromy  data 
$\left\{({\cal M}_1,{\cal M}_2,{\cal M}_3)\slash_\sim,
{\cal R}_1,{\cal R}_2,{\cal R}_3\right\}$ then this branch is 
unique modulo symmetries.

\vskip 0.5 cm
\noindent{\bf 3. Analytic continuation and rational solutions of 
the Painlev\'e VI equation.}
\vskip 0.3 cm

In Theorem 2.10, we parameterised branches of generic solutions of PVI
by triples of monodromy matrices. Following [DM], now we show how 
these parameters change with a change of the branch in the process of 
analytic continuation of the solutions along a path in 
$\overline\complessi\backslash\{0,1,\infty\}$. 
Recall that, as it follows from Theorem 2.5, any solution of the 
Schlesinger equations can be continued analytically from a point 
$(u_1^0,u_2^0,u_3^0)$ to another point $(u_1^1,u_2^1,u_3^1)$ along a path 
$$
(u_1(t),u_2(t),u_3(t))\in\complessi^3\backslash\{diags\},
\qquad 0\leq t\leq 1,
$$
where $\{diags\}=\{u_1,u_2,u_3|\, u_i=u_j,\,\hbox{for some}\, i\neq j\}$
and
$$
u_i(0)=u_i^0,\quad\hbox{and}\quad u_i(1)=u_i^1,
$$
provided that the end-points are not the poles of the solution. The result 
of the analytic continuation depends only on the homotopy class of the 
path in $\complessi^3\backslash\{diags\}$. Particularly, to find all the 
branches of a solution near a given point $u^0=(u_1^0,u_2^0,u_3^0)$ one 
has to compute the results of the analytic continuation along any homotopy 
class of closed loops in $\complessi^3\backslash\{diags\}$ with the 
beginning and the end at the point $u^0=(u_1^0,u_2^0,u_3^0)$.
Let 
$$
\beta\in\pi_1\left(\complessi^3\backslash\{diags\};\, u^0\right)
$$
be an arbitrary loop. Any solution of the Schlesinger equations near the point 
$u^0=(u_1^0,u_2^0,u_3^0)$, is uniquely determined up to \eqrefp{dig} 
by the monodromy matrices ${\cal M}_1$, ${\cal M}_2$ and 
${\cal M}_3$, computed with
respect to the basis of loops $\gamma_1$, $\gamma_2$, $\gamma_3$. 
Continuing analytically this solution along the loop $\beta$, we arrive at 
another branch of the same solution near $u^0$. This new branch is 
specified, according to Theorem 2.10, by some new monodromy matrices 
${\cal M}_1^\beta$, ${\cal M}_2^\beta$ and ${\cal M}_3^\beta$, 
computed in the same basis 
$\gamma_1$, $\gamma_2$, $\gamma_3$. We want to compute these 
new matrices for any loop 
$\beta\in\pi_1\left(\complessi^3\backslash\{diags\};u^0\right)$.
The fundamental group 
$\pi_1\left(\complessi^3\backslash\{diags\};\, u^0\right)$ is isomorphic 
to the pure (or unpermuted) braid group, $P_3$ with three strings 
(see [Bir]).

\proclaim Lemma 3.1. For the generators $\beta_1$, $\beta_2$, $\beta_3$
of the pure braid group $P_3$, ${\cal M}_i^\beta$ have the following form:
$$
{\cal M}_1^{\beta_1}=
{\cal M}_1{\cal M}_2{\cal M}_1 {\cal M}_2^{-1}{\cal M}_1^{-1},
\quad {\cal M}_2^{\beta_1}={\cal M}_1 {\cal M}_2 {\cal M}_1^{-1},
\quad {\cal M}_3^{\beta_1}= {\cal M}_3,\autoeqno{D2}
$$
$$
\eqalign{{\cal M}_1^{\beta_2} &={\cal  M}_1{\cal M}_3{\cal M}_1 
{\cal M}_3^{-1}{\cal M}_1^{-1},\qquad\quad
{\cal M}_3^{\beta_2}= {\cal M}_1 {\cal M}_3 {\cal M}_1^{-1},\cr
&{\cal M}_2^{\beta_2}= {\cal M}_1{\cal M}_3
{\cal M}_1^{-1}{\cal M}_3^{-1}{\cal M}_2
{\cal M}_3 {\cal M}_1 {\cal M}_3^{-1}{\cal M}_1^{-1}.\cr}
\autoeqno{D3}
$$
$$
{\cal M}_1^{\beta_3}={\cal M}_1
\quad {\cal M}_2^{\beta_3}={\cal M}_2 {\cal M}_3 {\cal M}_2 
{\cal M}_3^{-1} {\cal M}_2^{-1},
\quad {\cal M}_3^{\beta_3}= {\cal M}_2 {\cal M}_3 {\cal M}_2^{-1},
\autoeqno{D4}
$$
\noindent Proof. This lemma is proved in [DM].

\proclaim Lemma 3.2. If a solution of the Painlev\'e VI equation such 
that none of the corresponding monodromy matrices 
${\cal M}_1,{\cal M}_2,{\cal M}_3$ are multiples of the identity is 
rational then ${\cal M}_1,{\cal M}_2,{\cal M}_3$ are fixed 
points under the above action \eqrefp{D2}, 
\eqrefp{D3}, \eqrefp{D4} on the space of triples
$$
\left\{({\cal M}_1,{\cal M}_2,{\cal M}_3)\slash_\sim,\,
{\cal M}_3{\cal M}_2{\cal M}_1={\cal M}_\infty^{-1}\right\}
$$
the equivalence relation $\sim$ being defined in \eqrefp{amb2}.

\noindent{\bf Proof.} 
The action \eqrefp{D2}, \eqrefp{D3}, \eqrefp{D4} of the pure braid
group on the triples of monodromy matrices not only describes
the structure of the analytic continuation of the solutions of the 
Schlesinger equations \eqrefp{N10}, but also of the reduced ones 
\eqrefp{N13} and thus of the PVI equation. A necessary condition for
a solution to be rational is that it has only one branch. Since 
the action \eqrefp{D2}, \eqrefp{D3}, \eqrefp{D4} of the pure braid
group on the triples of monodromy matrices commutes with the
conjugation \eqrefp{amb2}, the triple of monodromy matrices 
${\cal M}_1,{\cal M}_2,{\cal M}_3$ corresponding to a rational
solution is unique up to \eqrefp{amb2}. {\hfill $\bigtriangleup$}

\proclaim Lemma 3.3. If a solution of PVI such that none of the corresponding 
monodromy matrices ${\cal M}_1,{\cal M}_2,{\cal M}_3$ are multiples of the 
identity is rational then the monodromy matrices all commute
$$
[{\cal M}_i,{\cal M}_j]=0,\qquad\forall\,i,j=1,2,3.\autoeqno{E1}
$$

\noindent{\bf Proof.} By Lemma 3.2, we have to impose
$$
\eqalign{
{L_\infty}_1^{-1} {\cal M}_1 {L_\infty}_1 &= {\cal M}_1^{\beta_1}=
{\cal M}_1{\cal M}_2{\cal M}_1{\cal M}_2^{-1}{\cal M}_1^{-1},\cr
{L_\infty}_1^{-1}{\cal M}_2 {L_\infty}_1 &= {\cal M}_2^{\beta_1}=
{\cal M}_1{\cal  M}_2 {\cal M}_1^{-1},\cr
{L_\infty}_1^{-1}{\cal M}_3 {L_\infty}_1 &= {\cal M}_3^{\beta_1}=
{\cal M}_3,\cr}\autoeqno{l1}
$$
$$
\eqalign{
{L_\infty}_2^{-1} {\cal M}_1 {L_\infty}_2 &= {\cal M}_1^{\beta_2}=
{\cal M}_1{\cal M}_3{\cal M}_1{\cal M}_3^{-1}{\cal M}_1^{-1},\cr
{L_\infty}_2^{-1}{\cal M}_2 {L_\infty}_2 &= {\cal M}_2^{\beta_2}=
{\cal M}_1{\cal  M}_3 {\cal M}_1^{-1}{\cal M}_3^{-1}{\cal M}_2
{\cal M}_3{\cal  M}_1 {\cal M}_3^{-1}{\cal M}_1^{-1},\cr
{L_\infty}_2^{-1}{\cal M}_3 {L_\infty}_2 &= {\cal M}_3^{\beta_2}=
{\cal M}_1 {\cal M}_3 {\cal M}_1^{-1},\cr}\autoeqno{l2}
$$
$$
\eqalign{
{L_\infty}_3^{-1} {\cal M}_1 {L_\infty}_3 &= 
{\cal M}_1^{\beta_3}= {\cal M}_1,\cr
{L_\infty}_3^{-1}{\cal M}_2 {L_\infty}_3 &= {\cal M}_2^{\beta_3}=
{\cal M}_2{\cal  M}_3 {\cal M}_2{\cal  M}_3^{-1} {\cal M}_2^{-1},\cr
{L_\infty}_3^{-1}{\cal M}_3 {L_\infty}_3 &= {\cal M}_3^{\beta_3}=
{\cal M}_2{\cal M}_3{\cal M}_2^{-1},\cr}\autoeqno{l3}
$$
for some suitable matrices ${L_\infty}_1$, ${L_\infty}_2$,
${L_\infty}_3$ that are diagonal for $\thi\not\in\interi$,
are in Jordan form for $\thi\in\interi$ and ${\cal R}_\infty\neq0$.
Then we have to distinguish two cases: 
i) ${\cal M}_\infty\neq\pm\ID$ is diagonal, 
ii) ${\cal M}_\infty\neq\pm\ID$ is in Jordan form.

i) In this case, the matrices ${L_\infty}_1,{L_\infty}_2,{L_\infty}_3$ 
are diagonal. If none of the matrices ${\cal M}_{1,2,3}$ is diagonal 
in the basis of ${\cal M}_\infty$ diagonal, then the above matrices 
${L_\infty}_1,{L_\infty}_3$ must be chosen to be multiples of the 
identity. Thus, from
$$
\eqalign{
{\cal M}_2&= {\cal M}_2^{\beta_1}=
{\cal M}_1{\cal  M}_2 {\cal M}_1^{-1},\cr
{\cal M}_3 &= {\cal M}_3^{\beta_3}=
{\cal M}_2 {\cal M}_3{\cal M}_2^{-1},\cr}
$$
it follows immediately that 
$$
[{\cal M}_2,{\cal M}_1]=0,\quad\hbox{and}\quad
[{\cal M}_2,{\cal M}_3]=0,
$$
thus
$$
{L_\infty}_2^{-1}{\cal M}_2{L_\infty}_2= {\cal M}_2^{\beta_2}=
{\cal M}_2,
$$
i.e. ${L_\infty}_2$ is the identity matrix as well and thus we obtain 
\eqrefp{E1}.
If one of the monodromy matrices ${\cal M}_{1,2,3}$ is diagonal in 
the basis of ${\cal M}_\infty$ diagonal, for example ${\cal M}_1$, 
then either ${L_\infty}_1$ is the identity matrix or ${\cal M}_3$ is diagonal.
If ${L_\infty}_1$ is the identity matrix, from \eqrefp{l1}
we have that ${\cal M}_2$ is diagonal as well, and being 
${\cal M}_\infty$ diagonal as well, ${\cal M}_3$ 
must be diagonal and we find again \eqrefp{E1}. If ${\cal M}_3$ and 
${\cal M}_1$ are both diagonal, since ${\cal M}_\infty$ is diagonal, 
${\cal M}_2$  must be diagonal too and we find again \eqrefp{E1}.
 
ii) Suppose that ${\cal M}_\infty$ is in Jordan form. Then the matrices 
${L_\infty}_1,{L_\infty}_2,{L_\infty}_3$ have the
form $\pmatrix{1& a_i\cr 0& 1\cr}$ for some constants $a_1,a_2,a_3$. 
If none of the
matrices ${\cal M}_1,{\cal M}_2,{\cal M}_3$ is in Jordan
form, all matrices ${L_\infty}_1,{L_\infty}_2,{L_\infty}_3$ are the 
identity matrix and we obtain \eqrefp{E1} as above. Suppose that one
of the monodromy matrices, for example ${\cal M}_1$ is in 
Jordan form. Then, either ${L_\infty}_1$ is the identity and 
$[{\cal M}_2,{\cal M}_1]=0$ or ${\cal M_3}$ is in
Jordan form. Reasoning as above we obtain that all matrices 
${\cal M}_1,{\cal M}_2,{\cal M}_3$ are in Jordan form and thus commute.
{\hfill $\bigtriangleup$}

\vskip 0.4 cm
\semiautosez{4}{4. Reducible monodromy groups.}

\proclaim Theorem 4.1. All solutions of PVI corresponding to reducible
monodromy groups are equivalent via birational canonical
transformations and symmetries to the following one-parameter family 
of solutions, realized for $\thi=-(\vartheta_1+\vartheta_2+\vartheta_3)$
$$
y={(1+\vartheta_1+\vartheta_2-x-\vartheta_2 x) u(x) -x(x-1) u_x(x)
\over (1+\vartheta_1+\vartheta_2+\vartheta_3) u(x)}\autoeqno{ric}
$$
where $u(x)=u_1(x)+\nu u_2(x)$, $u_1(x),u_2(x)$ being two linear independent
solutions of the following hypergeometric equation 
$$
x(1-x) u_{xx}(x)+\left[2+\vartheta_1+\vartheta_2-
(4+\vartheta_1+2\vartheta_2+\vartheta_3)x\right]
u_x(x)-(2+\vartheta_1+\vartheta_2+\vartheta_3)(\vartheta_2+1) u(x)=0.
\autoeqno{hyp}
$$

\noindent Proof. 
For reducible monodromy groups there exists a basis in which all
monodromy matrices are upper triangular. We can always perform
a change of basis in order that ${\cal M}_\infty$ has the form \eqrefp{N9} 
and all monodromy matrices have the form:
$$
{\cal M}_k=\pmatrix{\exp(\pi i\vartheta_k)&\star\cr 
0&\exp(-\pi i\vartheta_k)\cr}.
$$ 
It then follows, by the relation \eqrefp{N6}, that 
$\thi+\varepsilon_k\sum_k\vartheta_k=2 N$, $\varepsilon_k=\pm1$, $N\in\interi$.
By means of the birational canonical transformations and symmetries of
the PVI equation, we can always assume that $\varepsilon_k=+1$ and $N=0$.
Perform the following gauge transformation on the Fuchsian system
$$
\Phi\to\tilde\Phi=\Pi_{k=1}^3(\lambda-u_k)^{-\vartheta_k\over2}\Phi,
\qquad
{\cal A}_k\to\tilde{\cal A}_k={\cal A}_k-{\vartheta_k\over2}\ID.
$$ 
The new residue at infinity is
$$
\tilde{\cal A}_\infty=\pmatrix{0&0\cr
0&{-\thi+\sum\vartheta_k\over 2}\cr},
$$
and the new monodromy matrices are
$$
\tilde{\cal M}_k=\exp(-\pi i\vartheta_k){\cal M}_k
=\pmatrix{1&\star\cr 0&\exp(-2\pi i\vartheta_k)},\qquad k=1,2,3,
$$
and 
$$
\tilde{\cal M}_\infty=
\pmatrix{1&0\cr 0& 
\exp[\pi i(-\thi+\sum_k\vartheta_k)]\cr}\exp({\cal R}_\infty).
$$
Since all monodromy matrices have the first column given by 
$\pmatrix{1\cr0\cr}$, the new Fuchsian system admits a non-zero
single valued vector solution $\tilde Y$. This solution is analytic at $u_1$,
$u_2$ and $u_3$ because all residue matrices $\tilde{\cal A}_k$, $k=1,2,3$
have a zero eigenvalue. At infinity $\tilde Y$ is necessarily
constant. Thus, near each $u_j$, one has
$$
{\tilde{\cal A}_j\over\lambda-u_j} \tilde Y+{\cal O}(1)=0,
$$
that implies that $\tilde Y$ has the form $\pmatrix{a\cr0\cr}$, for
some $a\neq0$ and all residue matrices ${\cal A}_k$ are upper
triangular. Correspondingly $p=\sum_{k=1}^3{\vartheta_k\over(q-u_k)}$ 
and the solution $y(x)$ of PVI is given by \eqrefp{ric}.
{\hfill $\bigtriangleup$}
\vskip 0.2 cm 

\noindent{\bf Remark 4.2.} Observe that the appearance of the
hypergeometric equation in Theorem 4.1 is quite natural. In fact, 
when all the residue matrices ${\cal A}_k$ are upper-triangular,  
the Schlesinger equations, and thus the Painlev\'e VI equation, 
linearise.

\proclaim Lemma 4.3. The one-parameter family of classical
solutions \eqrefp{ric} contains at least one rational solution 
if and only if one of the values $\vartheta_1$, $\vartheta_2$, $\vartheta_3$, 
$\thi$ is an integer. 

\noindent Proof.\quad The one-parameter family of classical
solutions \eqrefp{ric} contains at least one algebraic solution if and only if
the corresponding Riccati equation 
$$
y_x(x)={1+\vartheta_1+\vartheta_2+\vartheta_3\over x(x-1)} y^2-
{1+\vartheta_1+\vartheta_2+\vartheta_1 x+\vartheta_3 x\over x(x-1)} y+
{\vartheta_1\over x-1}
$$
has at least one algebraic solution, i.e., if and only the corresponding
hypergeometric equation \eqrefp{hyp} is integrable in the sense
of differential Galois theory (see [Mor]). This happens if and only if and 
only if the parameters $\lambda$, $\mu$,$\nu$ of the hypergeometric equation 
belong to the Schwartz-Kimura table (see [Mor]). In particular
rational solutions are occur only when at least one of the numbers
$\mu-\lambda+\nu$, $\mu-\lambda-\nu$, $\mu+\lambda+\nu$, $\mu+\lambda-\nu$ 
is an odd integer. In our case $\lambda=-1-\vartheta_1-\vartheta_2$, 
$\mu=1+\vartheta_1+\vartheta_3$, $\nu=-1-\vartheta_3-\vartheta_2$ so the
one-parameter family of classical solutions \eqrefp{ric} contains at 
least one rational solution if and only if one of the values $\vartheta_1$, 
$\vartheta_2$, $\vartheta_3$, $\thi$ is an integer. 
{\hfill $\bigtriangleup$}

\vskip 0.2 cm
To conclude the proof of the main theorem, we use the following lemma

\proclaim Lemma 4.4. All solutions of PVI such that one or more
monodromy matrices ${\cal M}_1,{\cal M}_2, {\cal M}_3,{\cal M}_\infty$
is a multiple of the identity are either degenerate or are equivalent 
via birational canonical transformations to the one-parameter family 
(\eqref{ric}) of Theorem 4.1.

\noindent Proof.\quad We omit the proof of this Lemma that can be found 
for example in [Ma1].

\vskip 0.2 cm
The classification is concluded observing that if one or more of the 
monodromy matrices is a multiple of the identity, then the corresponding
solution is equivalent via birational canonical transformations to a solution
with reducible monodromy groups. If none of the monodromy matrices is a
multiple of the identity, then in order to have a rational solution the
monodromy group must be Abelian. All Abelian $2\times2$ 
monodromy groups are reducible. Thus all rational non-degenerate solutions
of the PVI equation belong to the one parameter family of classical
solutions given in Theorem 4.1, up to birational canonical transformations. 
Then we can use Lemma 4.3 to classify them.

\vskip 0.5 cm

\noindent{\bf BIBLIOGRAPHY }
\vskip 0.3 cm
\item{[Ai]}
H. Airault, Rational solutions of Painlev\'e equations, {\it Stud. Appl. 
Math.}\/ {\bf 61}, (1979) 31--53.
\vskip 0.2 cm

\item{[AMM]}
D.A. Albrecht, E.L. Mansfield, A.E. Milne, Algoritms for special integrals 
of ordinary differential equations, {\it J. Phys. A: Math. Gen.}\/ {\bf 29}
(1996) 973--991.
\vskip 0.2 cm

\item{[Bir]}
J.S. Birman, {\it Braids, Links, and Mapping Class Groups,}\/ Ann. Math. 
Stud. Princeton University (1975).
\vskip 0.2 cm

\item{[Dub]} 
B. Dubrovin, Painlev\'e Transcendents in Two-Dimensional Topological
Field Theory, {\it The Painlev\'e Property One Century Later,}\/
Robert Conte editor, CRM Series in Mathematical Physics, (1999).
\vskip 0.2 cm

\item{[DM]} 
B. Dubrovin, M.Mazzocco, Monodromy of certain Painlev\'e VI
transcendents and Reflection Groups, {\it Invent. Math.,}\/
{\bf 141} (2000) 55--147.
\vskip 0.2 cm

\item{[FlN]}
H. Flashka and A.C. Newell, Monodromy and Spectrum Preserving 
Deformations, {\it Comm. Math. Phys.}\/  {\bf 76} (1980) 67--116.
\vskip 0.2 cm

\item{[Fuchs]}
R. Fuchs, \"Uber Lineare Homogene Differentialgleichungen Zweiter 
Ordnung mit im drei im Endrichen Gelegene Wesentlich Singul\"aren 
Stellen, {\it Math. Ann.}\/ {\bf 63} (1907) 301--321.
\vskip 0.2 cm

\item{[Gam]}   
B. Gambier, Sur les Equations Differentielles du Second Ordre et du 
Primier Degr\`e dont l'Integrale est a Points Critiques Fixes, {\it Acta 
Math.}\/  {\bf 33}, (1910) 1--55.
\vskip 0.2 cm

\item{[Gar]}
R. Garnier, Sur des Equations Differentielles du Troisieme Ordre dont 
l'Integrale Generale est Uniforme et sur una Classe d'Equations nouvelles 
d'Ordre Superieur dont  l'Integrale Generale a ses Points Critiques Fixes,
{\it Ann. Sci. Ecole Norm. Sup.}\/  {\bf 3} No.29 (1912) 1--126.
\vskip 0.2 cm

\item{[Gr]}
V.I. Gromak, Single-parameter families of solutions of Painlev\'e
equations, {\it Diff. Eqns.}\/ {\bf 14} (1978), 1510--1513.
\vskip 0.2 cm

\item{[GL]}
V.I. Gromak, N.A. Lukashevich, Special classes of solutions of Painlev\'e 
equations, {\it Diff. Eqns.}\/ {\bf 18} (1982) 317--326.
\vskip 0.2 cm

\item{[Hit]}
N. Hitchin, Hypercomplex Manifolds and the Space of Framings, {\it The 
Geometric Universe,}\/ Oxford Univ. Press, Oxford, 1998, 9--30. 
\vskip 0.2 cm

\item{[Ince]}
E.L. Ince, {\it Ordinary Differential Equations,}\/
Dover Publications, New York (1956).
\vskip 0.2 cm

\item{[ItN]} 
A.R. Its and V.Yu. Novokshenov, The Isomonodromic Deformation Method 
in the Theory of Painlev\'e Equations, {\it Springer Lect. Notes Math.}\/ 
{\bf 1191} (1986).
\vskip 0.2 cm

\item{[JMU]}
M. Jimbo, T. Miwa and K. Ueno, Monodromy Preserving Deformation of 
the Linear Ordinary Differential Equations with Rational Coefficients I, 
II, III, {\it Physica D,} {\bf 2}  (1981), no. 2, 306--352, 
{\bf 2} (1981), no. 3, 407--448,  {\bf 4} (1981/82), no. 1, 26--46.
\vskip 0.2 cm

\item{[Luk]}
N.A. Lukashevich, Elementary solutions of certain Painlev\'e equations, 
{\it Differ. Uravn.}\/ {\bf 1} (1965) 731--735.
\vskip 0.2 cm

\item{[Mal]}
B. Malgrange, Sur les Deformations Isomonodromiques I, Singularit\'es 
R\'eguli\`eres, Seminaire de l'Ecole Normale Superieure 1979--1982, 
{\it Progress in Mathematics}\/ {\bf 37}, Birkh\"auser, Boston (1983) 401--426.
\vskip 0.2 cm

\item{[Ma]}
M. Mazzocco, Picard and Chazy Solutions to PVI Equation, {\it Math. Ann.,}\/ 
{\bf 321} (2001), 157--195.
\vskip 0.2 cm

\item{[Ma1]}
M. Mazzocco, The geometry of the classical solutions of the Garnier systems,
{\it I.M.R.N.,}\/ {\bf 12} (2002) 613--646.
\vskip 0.2 cm

\item{[Miwa]}
T. Miwa, Painlev\'e Property of Monodromy Preserving Deformation 
Equations and the Analyticity of $\tau$-function, {\it Publ. RIMS, Kyoto 
Univ.}\/ {\bf 17} (1981) 703--721.
\vskip 0.2 cm

\item{[Mor]}
J.J. Morales, {\it Differential Galois theory and non-integrability of
Hamiltonian systems}\/ Progress in Mathematics, {\bf 179}, (1999). 
\vskip 0.2 cm

\item{[Ok]}
K. Okamoto, Studies on the Painlev\'e Equations I, Sixth Painlev\'e 
Equation, {\it Ann. Mat. Pura Appl.}\/ {\bf 146} (1987) 337--381.
\vskip 0.2 cm

\item{[Pain]}
P. Painlev\'e, Sur les Equations Differentielles du Second Ordre et
d'Ordre Superieur, dont l'Interable Generale est Uniforme, {\it Acta Math.}\/ 
{\bf 25} (1902) 1--86.
\vskip 0.2 cm

\item{[Sch]}
L. Schlesinger, \"Uber eine Klasse von Differentsial System 
Beliebliger Ordnung mit Festen Kritischer Punkten, {\it J. fur Math.}\/
{\bf 141} (1912), 96--145.
\vskip 0.2 cm

\item{[Schw]}
H.A. Schwartz, \"Uber Diejenigen F\"alle in Welchen die Gaussische 
Hypergeometrische Reihe einer Algebraische Funktion iheres vierten 
Elementes Darstellit, {\it Crelle J.}\/ {\bf 75} (1873) 292--335.
\vskip 0.2 cm

\item{[Sib]}
Y. Sibuya, {\it Linear Differential Equations in the Complex Domain: Problems 
of Analytic Continuation,}\/ AMS TMM {\bf 82} (1990).
\vskip 0.2 cm

\item{[Um]}
H. Umemura, 
Special polynomials associated with the Painlev\'e equations I, to appear in
{\it Painlev\'e transcendents,}\/  CRM Montreal, Canada (1996). 
\vskip 0.2 cm

\item{[Um1]}
H. Umemura,  Irreducebility of the First Differential Equation of 
Painlev\'e, {\it Nagoya Math. J.,}\/ {\bf 117} (1990) 231--252.
\vskip 0.2 cm

\item{[Um2]}
H. Umemura, Second proof of the irreducebility of the First
Differential 
Equation of Painlev\'e, {\it Nagoya Math. J.,}\/ {\bf 117} (1990) 125--171.
\vskip 0.2 cm

\item{[Wat]}
H. Watanabe, Birational Canonical Transformations and Classical Solutions 
of the Sixth Painlev\'e Equation, {\it Ann. Scuola
Norm. Sup. Pisa Cl. Sci.}\/  {\bf 27} (1998), no. 3-4, 379--425
(1999).
\vskip 0.2 cm

\bye